\documentclass{aa}
\usepackage{txfonts}
\usepackage{graphicx}
\usepackage{natbib}
\usepackage{amssymb}
\usepackage{latexsym}
\bibpunct{(}{)}{;}{a}{}{,}

\def\xmm{{\it XMM-Newton}}
\def\intgr{{\it INTEGRAL}}

\def\cra{{\it Chandra}}

\begin{document}
\title{INTEGRAL probes the morphology of the Crab nebula in hard X-rays/soft $\gamma$-rays}
\author{D. Eckert\inst{1,2}, V. Savchenko\inst{1}, N. Produit\inst{1} \& C. Ferrigno\inst{1,3}}
\offprints{Dominique Eckert (Occhialini fellow), \email{eckert@iasf-milano.inaf.it}}

\institute{$^1$ISDC Data Centre for Astrophysics, Geneva Observatory, University of Geneva, 16, ch. d'Ecogia, CH-1290 Versoix, Switzerland\\
$^2$INAF/IASF Milano, Via E. Bassini 15, 20133 Milano, Italy\\
$^3$IAAT, Abt.\ Astronomie, Universit\"at T\"ubingen, Sand 1, 72076 T\"ubingen, Germany}

\date{Received May 29, 2009/ Accepted October 08, 2009}

\abstract{}{We use the IBIS/ISGRI telescope on-board \emph{INTEGRAL} to measure the position of the centroid of the 20-200 keV emission from the Crab region.}{We find that the astrometry of the IBIS telescope is affected by the temperature of the IBIS mask during the observation. After correcting for this effect, we show that the systematic errors on the astrometry of the telescope are of the order of 0.5 arcsec. In the case of the Crab nebula and several other bright sources, the very large number of photons renders the level of statistical uncertainty in the centroid smaller or comparable to this value.}{We find that the centroid of the Crab nebula in hard X-rays (20-40 keV) is shifted by 8.0 arcsec with respect to the Crab pulsar in the direction of the X-ray centroid of the nebula. A similar shift is also found at higher energies (40-100 and 100-200 keV). We observe a trend of decreasing shift with energy, which can be explained by an increase in the pulsed fraction. To differentiate between the contribution of the pulsar and the nebula, we divide our data into an on-pulse and off-pulse sample. Surprisingly, the nebular emission (i.e., off-pulse) is located significantly away from the X-ray centroid of the nebula.}{In all 3 energy bands (20-40, 40-100 and 100-200 keV), we find that the centroid of the nebula is significantly offset from the predicted position. We interpret this shift in terms of a cut-off in the electron spectrum in the outer regions of the nebula, which is probably the origin of the observed spectral break around 100 keV. From a simple spherically-symmetric model for the nebula, we estimate that the electrons in the external regions of the torus ($d\sim0.35$ pc from the pulsar) reach a maximal energy slightly below $10^{14}$ eV.}

\authorrunning{Eckert D. et al.}
\titlerunning{The centroid of the Crab nebula in hard X-rays/soft $\gamma$-rays}

\maketitle

\section{Introduction}

The Crab nebula is the brightest astrophysical source in the $\gamma$-ray sky ($E>30$ keV). It is the prototypical pulsar-wind nebula (PWN, see e.g., \citet{gaensler} for a review), where material ejected from the central pulsar at different epochs interacts, producing strong shocks that accelerate electrons up to energies $\sim10^{15}$ eV \citep{kennel}. The synchrotron emission produced by the large population of non-thermal electrons in a magnetic field $B\sim0.3$ mG \citep{marsden} is detected over more than 10 orders of magnitude in the electromagnetic spectrum from radio to $\gamma$-rays \citep{atoyan,volpi,hesterreview}. 

In soft X-rays, the size of the nebula is $\sim$2 arcmin \citep{weisskopf}. It consists of a characteristic jet+torus structure, probably corresponding to relativistic outflows along the rotation axis and the equator of the pulsar. High-resolution \emph{Chandra} observations also revealed the presence of an inner ring, probably associated with the conversion of the relativistic pulsar wind into a synchrotron-emitting plasma. The measured photon index decreases with radius from $\Gamma\sim1.6$ around the pulsar down to $\Gamma\sim3.0$ in the outer parts of the X-ray image \citep{mori,willingale}, which implies that the most energetic electrons are constantly injected by the pulsar in the inner regions of the nebula. The steeper photon index in the outer regions can be explained by synchrotron cooling. This interpretation is supported by the observation of rapid X-ray variability in the inner ring \citep{hester}, which is probably due to the presence of strong quasi-stationary shocks. The pulsar itself exhibits a hard spectrum $\Gamma\sim1.6$ and a period $P\sim30$ msec with a double-peak profile.

In the hard X-ray/soft $\gamma$-ray range, early HEAO A-4 observations inferred a photon index of $\Gamma=2.15$ \citep{jung}. Around 100 keV a spectral break was detected, and above this energy a photon index of $\sim$2.5 was found. This measurement agrees with the steeper spectral indices measured at higher energies by COMPTEL \citep{strong} and EGRET \citep{nolan}. Therefore, observations of the Crab complex around the break energy by modern experiments are important to constrain particle acceleration models. Recently, a polarized $\gamma$-ray signal from the Crab was measured by the SPI \citep{dean} and IBIS \citep{forot} instruments on \intgr. The polarization was found to be co-aligned with the spin orientation, thus indicating a possible association of the polarized signal with the inner jet. The broad-band coverage of \intgr\ also permitted a study of the evolution of the pulse profiles as a function of energy \citep{mineo}. A phase-revolved analysis detected a significantly harder signal ($\Gamma\sim1.8$) during the interpulse phase compared to the peaks ($\Gamma\sim2.2$). However, the results presented there only considered the emission from the pulsar, and no information was given about the spectrum of the nebula.

In this paper, we use the coded-aperture IBIS telescope \citep{Ubeetal03} on-board \intgr\ \citep{Winetal03} to measure the position of the hard X-ray/soft $\gamma$-ray centroid of the Crab pulsar/nebula complex. In Sect. \ref{astrom}, we analyse the astrometry of the IBIS telescope and show that the point-source localization accuracy of the instrument depends on the temperature of the mask. We correct for this effect and assess the level of systematic uncertainties in the astrometry of IBIS. In Sect. \ref{results}, we report precise measurements of the position of the Crab pulsar/nebula and present our results. The implications of these findings are discussed in Sect. \ref{disc}.

\section{Astrometry of the IBIS telescope}
\label{astrom}

\subsection{IBIS misalignment correction}
\label{thermal}

The IBIS telescope, and in particular its low-energy detector layer ISGRI \citep{Lebetal03}, operates in the hard X-ray/soft $\gamma$-ray range (15-400 keV) and has an angular resolution of 12 arcmin FWHM. While this is much larger than the angular extent of the Crab nebula, the point source location accuracy of the instrument scales with the inverse of the detection significance of the source

\begin{equation}
\sigma_{PSF}\sim\frac{\mbox{FWHM}}{R(S/N)},
\end{equation}

\noindent where $R=2.43$ is the ratio of the size of mask to detector pixels \citep{goldwurm}, so in principle for a very bright source such as the Crab nebula ($S/N\sim3,000$ in the 20-40 keV band) IBIS/ISGRI can measure the position of the peak with an accuracy of $\sim0.1$ arcsec. However, it was noticed early in the mission \citep{Waletal03} that the axis of the star trackers was offset with respect to the spacecraft axis, introducing a systematic error in the positions measured by the telescope. To correct for this effect, a misalignment matrix performing a rotation between the star tracker and spacecraft axes was introduced, which allowed for the localization of astrophysical sources with a reasonable accuracy. However, for very bright sources (e.g., Crab, Cyg X-1, Sco X-1) a systematic shift of $\sim$ 7 arcsec with respect to the true source position was present, and thus for these sources it was not possible to measure the position with an accuracy greater than this value.

\begin{figure}
\resizebox{\hsize}{!}{\includegraphics{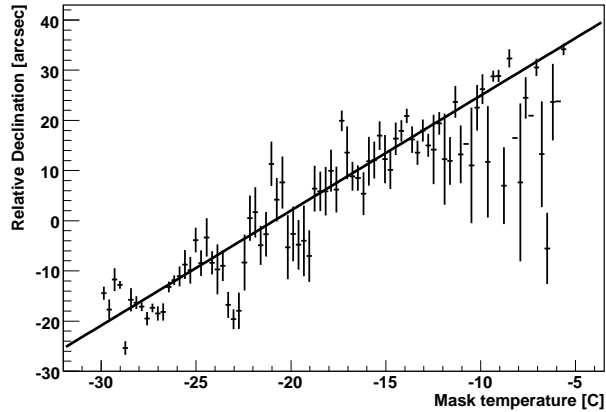}}
\caption{Relative Declination of Cyg X-1 with respect to the catalog position (in arcsec) as a function of the IBIS mask temperature (in $^\circ$C). The black line shows a linear fit to the data.}
\label{tmask}
\end{figure}

In the context of the development of the Offline Scientific Analysis \citep[OSA,][]{Couetal03} version 8.0, we investigated this effect analyzing a large sample of Cyg X-1 observations ($\sim$ 900 individual pointings (Science Windows, ScWs) each of typical duration 2-3 ksec) and computed the alignment of the instrument from scratch. In particular, we investigated the effects of the temperature of the IBIS mask on the source localization accuracy. Indeed, the dilatation of the IBIS mask and its supporting structure produced by the relative position of the Sun might have an influence on the alignment of the instrument. Figure \ref{tmask} shows the relative declination of Cyg X-1 with respect to the catalog position (in arcsec) as a function of the mask temperature. A linear trend can clearly be seen in the figure, and hence it seems that the mask temperature has a significant effect on the source localization accuracy of IBIS/ISGRI.

To correct for this effect, we introduced a linear dependence of the three rotation angles (i.e., Euler angles around the three spacecraft axes) on the mask temperature, and fitted the angles using a least squares method to achieve the closest possible agreement between the observed and catalog positions. We found that only one angle depends significantly on temperature, and therefore only one additional parameter is needed to correct for the temperature effect. The dependence of this angle on the temperature was measured accurately by the fit. Finally, we implemented these results into the scientific analysis software.

\subsection{Test of the IBIS astrometry on several point sources}
\label{syst}

To validate our method and test it on well-known cases, we selected a large sample of data comprising 2,017 individual pointings and containing at least one of the 3 brightest point sources in the ISGRI band (Sco X-1, GRS 1915+105, and V0332+53 during its 2005 outburst, but excluding Cyg X-1 since it was used as calibration source) and measured the position of the sources in each ScW (20-40 keV band). 

\begin{figure}
\centerline{\vbox{\includegraphics[width=\hsize]{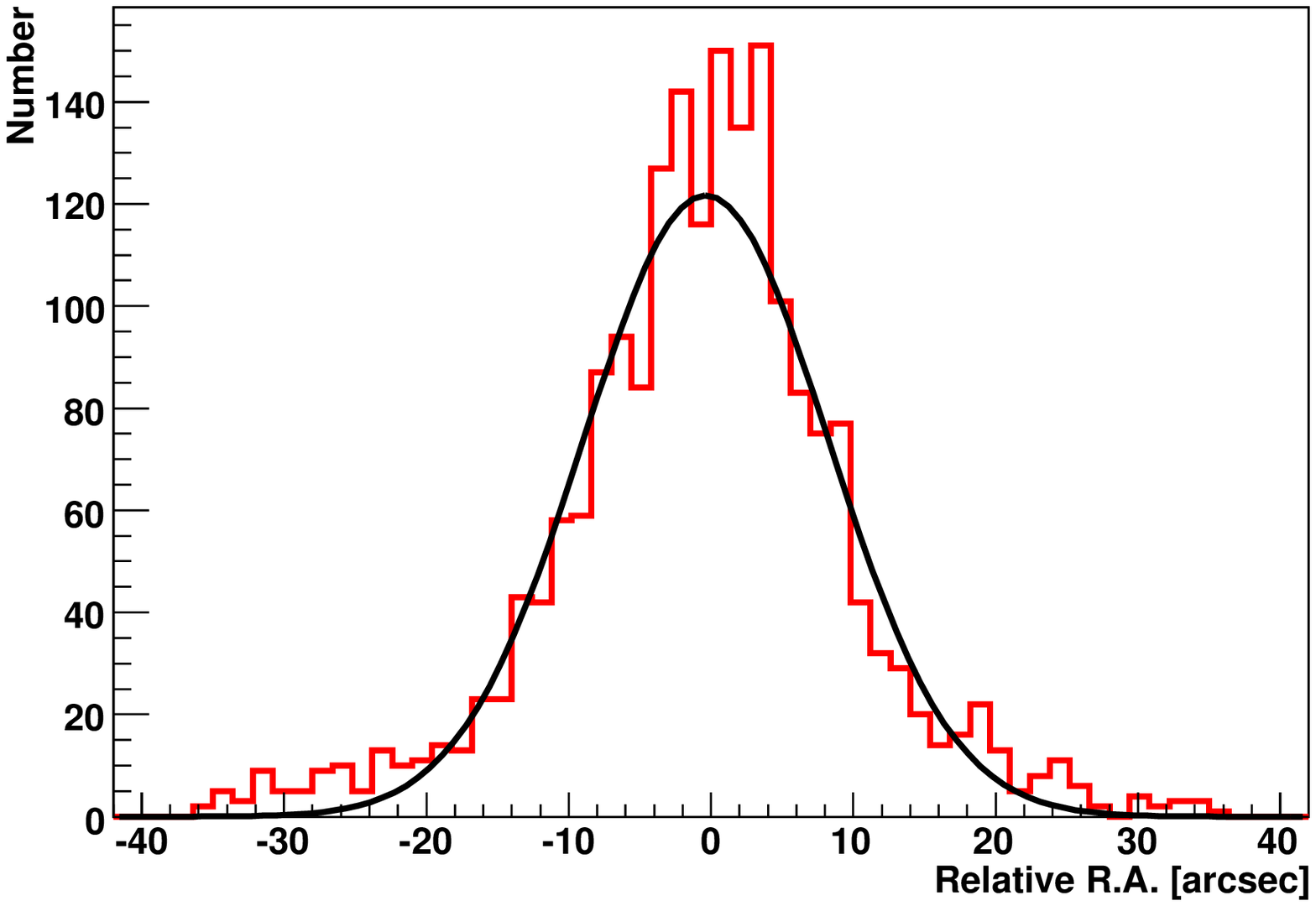}
\includegraphics[width=\hsize]{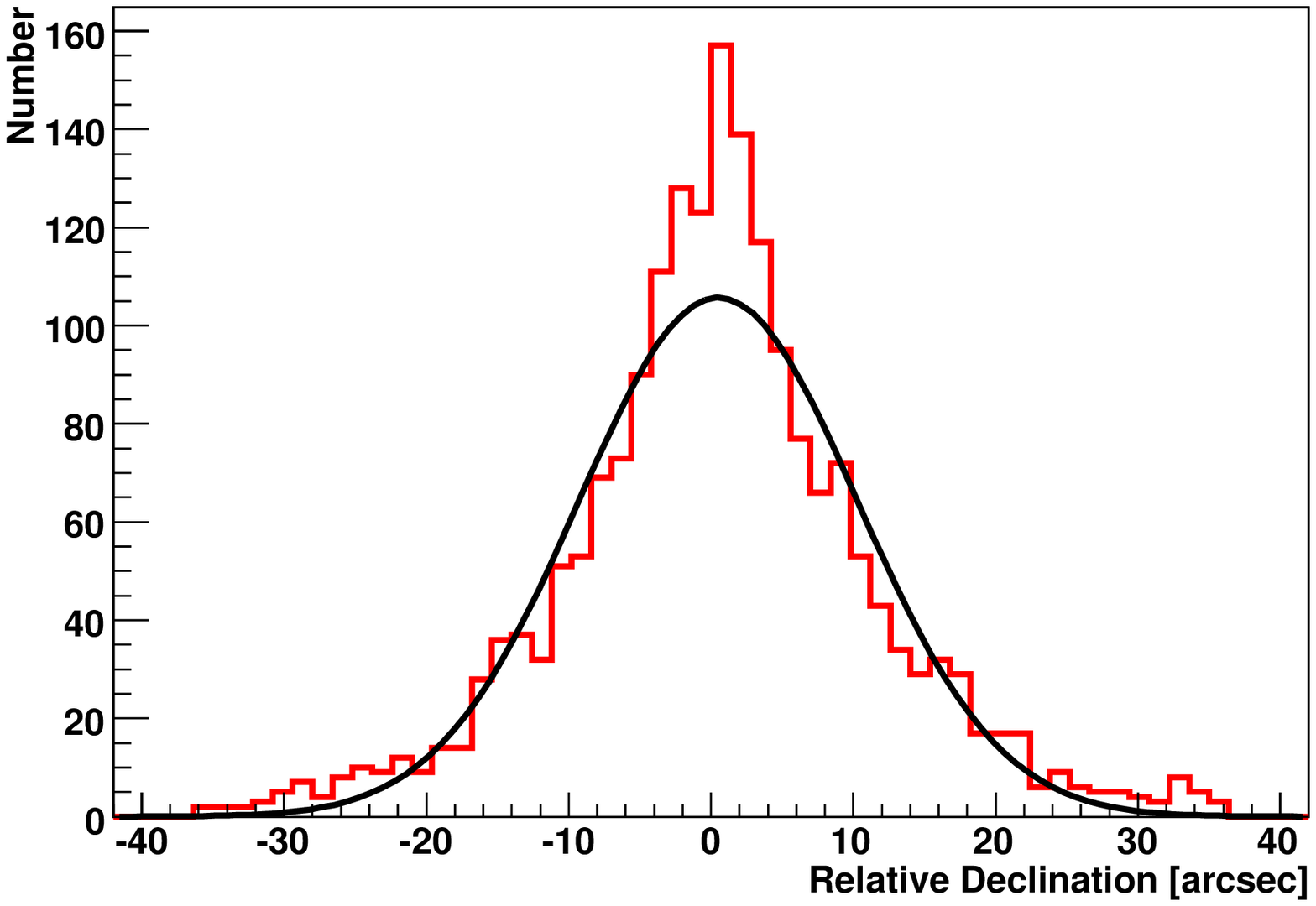}}}
\caption{Test of the IBIS/ISGRI astrometry on 3 of the brightest point sources in the 20-40 keV band (Sco X-1, GRS 1915+105 and V0332+53). The plots show the relative shift in RA (top) and dec (bottom) measured by the telescope with respect to the catalog position. Both histograms are fitted by a Gaussian profile (black curve).}
\label{4sources}
\end{figure}

We restricted our observations to pointings with low off-axis angle ($<8^\circ$), since the width of the IBIS PSF increases significantly for larger angles \citep{Groetal03}. Using these data, we constructed histograms of the relative shift in RA and Dec of the measured source position with respect to the catalog one. The resulting plots are presented in Fig. \ref{4sources}. Both distributions are well-represented by a Gaussian profile with a similar standard deviation $\sigma=8.2\pm0.1$ arcsec (RA) and $\sigma=8.1\pm0.1$ arcsec (Dec). The mean of the distribution is found to be $\mu=-0.35\pm0.21$ arcsec (RA) and $\mu=0.51\pm0.23$ arcsec (Dec). Similar shifts (although less accurate statistically) are obtained when each source is analysed individually. Based on these results, from now on we adopt the value of 0.5 arcsec as an estimate of the systematic uncertainties in the astrometry of the IBIS telescope.

The reader might argue that the systematic errors adopted here are smaller than the uncertainties in the attitude reconstruction of the telescope \citep[3 arcsec,][]{Waletal03}. However, it should be noted that the uncertainties in the attitude reconstruction affect each pointing individually in a non-preferential way. Given that we are analysing a large sample of pointings, this introduces a hard limit to the \emph{width} of the distribution ($\sim8"\gg3"$), but does not affect the \emph{mean} of the distribution.

In addition, one might also wonder whether the alignment of the instrument depends on energy, since the alignment was performed using low-energy data (20-40 keV) where count rate and signal-to-noise ratio are maximal. However, we note that the misalignment between the instrument and the star trackers is caused only by geometrical and thermal effects, which do not depend on the energy of the incoming photons. Therefore, for the remainder of the paper we use the same alignment and systematic errors in all energy bands.

\section{Results}
\label{results}

\subsection{Measuring the centroid of the Crab nebula}
\label{centroid}

As shown in Sect. \ref{syst}, we are now confident that we can measure the position of sources with a level of systematic uncertainties of about 0.5 arcsec. Since the pulsed fraction is less than 20\% in the \intgr\ energy range, while it is $\sim$10\% in the 1-10 keV, the hard X-ray emission is dominated by the nebula and thus IBIS/ISGRI can measure its position with excellent accuracy. We selected a data set from the ISDC archive\footnote{\texttt{http://isdc.unige.ch/?Data+browse}} only for pointings for which the offset angle of the Crab with respect to the spacecraft axis was at most 8.0 degrees, and we filtered out the observations affected by Solar flares. Overall, our sample comprises 1,141 individual ScWs of the Crab nebula, of a total effective exposure time of $\sim$1.8 Msec.

\begin{figure}
\resizebox{\hsize}{!}{\includegraphics{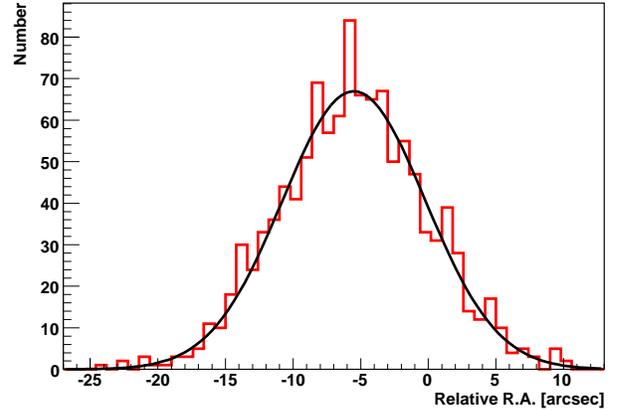}}
\caption{Relative right ascension (R.A.) of the best-fit position of Crab in the 20-40 keV band with respect to the position of the Crab pulsar. The black solid line shows a Gaussian fit to the data.}
\label{crabtot}
\end{figure}

The analysis was performed with the standard OSA software v.8.0, including the correction of thermal effects presented in Sect. \ref{thermal}. For each pointing, an image was created in the 20-40, 40-100, and 100-200 keV bands, and the position of the Crab nebula was fitted by a 2-dimensional Gaussian with the width of the IBIS PSF. The best-fit position was then collected, and histograms were computed. Figure \ref{crabtot} shows the distribution of the Crab best-fit position (in RA) in the 20-40 keV band with respect to the Crab pulsar. The distribution is obviously shifted from the Crab pulsar by $\sim$5.5 arcsec to the west. A similar shift is also found in declination by $\Delta$Dec$\sim$6.3 arcsec towards the north. We then fitted the distributions by a Gaussian profile to extract the best-fit positions (see Table \ref{tabpos}). In the 20-40 keV band, the centroid of the Crab nebula is found to be at $\mbox{RA}=83.6317 \pm 0.1"_{\mbox{\tiny{stat}}}\pm0.5"_{\mbox{\tiny{syst}}}$, dec $=22.0162\pm 0.1"_{\mbox{\tiny{stat}}}\pm0.5"_{\mbox{\tiny{syst}}}$. This corresponds to an angular distance to the Crab pulsar of $8.0\pm0.1_{\mbox{\tiny{stat}}}\pm0.5_{\mbox{\tiny{syst}}}$ arcsec. The fitted position is away from the Crab pulsar by more than $13\sigma$.

Given that the timing resolution of ISGRI \citep[$90\,\mu$s,][]{Lebetal03} is sufficient to resolve the on- and off-pulse emission from the pulsar \citep[$P=30$ ms,][]{tennant}, we extracted phase-resolved images of the Crab nebula for both the on- and off-pulse phases (see Table \ref{tabpos}). For convenience, the 2 phase bins were chosen to be equally-long, the on-pulse bin comprising both peaks and the interpulse \citep[see ][]{mineo}. More specifically, the on-pulse bin includes phases 0.95-0.45 (see Fig. \ref{folded}). In the 40-100 and 100-200 keV bands, we were also able to measure the centroid of the total emission with sufficient accuracy (see Table \ref{tabpos}).

\begin{table}
\begin{tabular}{lcccc}
\hline
\, & RA [$^\circ$] & Dec [$^\circ$] & $\sigma_{\mbox{\tiny{stat+syst}}}$ ["] & Shift ["]\\
\hline
\hline
Total, 20-40 keV & 83.6317 & 22.0162 & 0.6 & 8.0\\
Total, 40-100 keV & 83.6319 & 22.0159 & 0.7 & 6.7\\
Total, 100-200 keV & 83.6321 & 22.0157 & 1.0 & 5.8\\
Off-pulse, 20-40 keV & 83.6315 & 22.0164 & 0.7 & 9.0\\
On-pulse, 20-40 keV & 83.6319 & 22.0160 & 0.7 & 7.0\\
Off-pulse, 40-100 keV & 83.6316 & 22.0161 & 0.8 & 8.0\\
On-pulse, 40-100 keV & 83.6322 & 22.0158 & 0.7 & 5.9\\
Off-pulse, 100-200 keV & 83.6316 & 22.0162 & 1.2 & 8.2\\
On-pulse, 100-200 keV & 83.6323 & 22.0155 & 1.1 & 4.8\\
\hline
\end{tabular}
\caption{IBIS/ISGRI best-fit positions of the Crab in 3 energy bands. The phase-resolved positions (on- and off- pulse) are also reported. The $1\sigma$ error on the position ($3^{rd}$ column) is the sum of the statistical and systematic uncertainties, where the systematic error is fixed to be 0.5" (see Sect. \ref{syst}). The last column gives the angular distance between the measured position and the pulsar (in arcsec).}
\label{tabpos}
\end{table}

\subsection{Comparison with soft X-rays}
\label{interp}

\begin{figure}
\resizebox{\hsize}{!}{\includegraphics{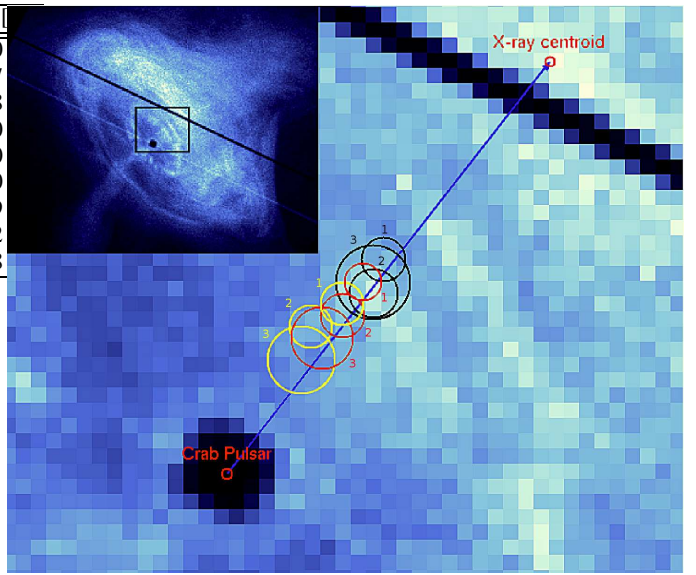}}
\caption{\emph{Chandra}/ACIS-S image of the central region of the Crab nebula, with the best-fit ISGRI positions in the 20-40 (band 1), 40-100 (2) and 100-200 keV (3) bands overlayed. The red circles show the best-fit positions and $1\sigma$ error circles of the total emission, while the black (yellow) circles show the measured positions of the off-pulse (on-pulse) phase bins. The position of the Crab pulsar and of the X-ray centroid of the nebula are also displayed. The size of the blue arrow is 17.3". For comparison, the inset shows a global view of the nebula. The black rectangle represents the size of the region displayed here.}
\label{chandra}
\end{figure}

To visualize the best-fit positions presented in Sect. \ref{centroid}, we used an archival high-resolution \cra\ observation, and extracted images in the full band (0.5-10.0 keV) as well as in a soft (0.5-2.0 keV) and a hard band (2.0-10.0 keV). In the \cra\ observation, the very strong pile-up effect at the position of the pulsar leads to the rejection of the events \citep{weisskopf}, which allows us to analyse the nebula individually. To estimate the centroid of the nebular emission, we convolved the full-band \cra\ image with a very broad Gaussian and fitted the resulting image with a 2-dimensional Gaussian. The resulting best-fit position for the centroid lies 17.3" away from the pulsar, in the torus and along the axis of the jet. Figure \ref{chandra} shows the best-fit positions with $1\sigma$ error circles shown in Tab. \ref{tabpos} for the total emission and the on- and off-pulse cases, superimposed on the \emph{Chandra} image of the region. We can see that all the measured positions lie less than $1\sigma$ away from the line joining the Crab pulsar to the X-ray centroid of the nebula, which corresponds roughly to the jet line from the pulsar. This indicates that the measured positions are a combination of the pulsar position with the position of the centroid of the nebula.

Interestingly, we can see in Fig. \ref{chandra} that all the measurements are found to be significantly offset from the centroid of the X-ray emission. Given that the pulsed emission accounts only for $\sim15\%$ of the flux, this indicates that the centroid of the emission from the nebula, i.e., its morphology, varies relative to the soft X-ray band. This statement is confirmed by the measurement of the off-pulse emission (where the signal is completely dominated by the nebula), which is also found to be significantly offset from the X-ray centroid of the nebula.

\subsection{Evolution of the pulsed fraction}
\label{pulsfrac}

To compare the results described above with the expected values, we investigated the evolution of the pulsed fraction with energy. Indeed, given that the measured centroid is the sum of the point-like signal from the pulsar and of the unresolved nebular emission, any evolution of the pulsed fraction with energy will influence the position of the centroid. 

\begin{figure}
\resizebox{\hsize}{!}{\includegraphics{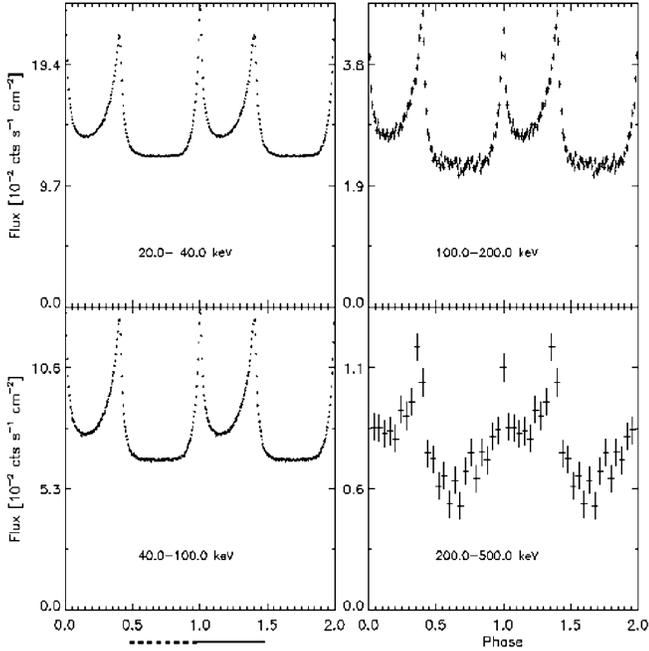}}
\caption{Background-subtracted pulse profiles of the Crab in the 20-40, 40-100, 100-200 and 200-500 keV bands. In the bottom left corner, the dashed line shows the off-pulse phase bin (0.45-0.95) while the solid line shows the on-pulse phase bin.}
\label{folded}
\end{figure}

We performed a phase-resolved analysis using the ephemeris of \citet{lyne}\footnote{\texttt{http://www.jb.man.ac.uk/$\sim$pulsar/crab.html}} and exploiting the method of \citet{segreto}. We collected all the Crab observations within an off-axis angle of 8 degrees (exposure of about 1.8~Ms) and produced a folded light curve with 200 phase bins, using energy bins of width 0.4787\,keV. Figure \ref{folded} shows the resulting background-subtracted pulse profiles in the 20-40, 40-100, 100-200, and 200-500 keV bands. The characteristic double-peak profile is clearly seen. In addition, the figure shows that during the off-pulse phase (0.45-0.95) the measured flux is nearly constant, which indicates that the pulsar contributes very little to the emission during the off-pulse phase.

\begin{figure}
\resizebox{\hsize}{!}{\includegraphics{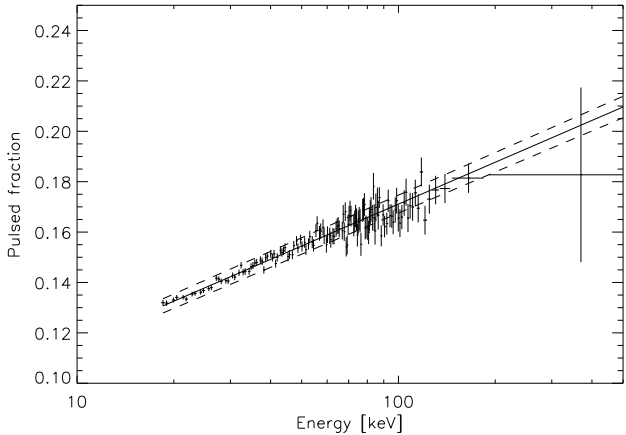}}
\caption{Pulsed fraction of the Crab pulsar/nebula complex as a function of energy. The solid line shows a linear fit to the data, with $1\sigma$ error given by the dashed lines.}
\label{pulsed}
\end{figure}

We then determined the pulsed fraction as a function of energy following the definition $(F_{\mbox{\tiny{on}}} - F_{\mbox{\tiny{off}}})/(F_{\mbox{\tiny{on}}} + F_{\mbox{\tiny{off}}})$, where $F_{\mbox{\tiny{on}}}$ is the total fluence during the pulse phase bins (phase 0.95-0.45), and $F_{\mbox{\tiny{off}}}$ is the fluence during the off-pulse bins (phase 0.6-0.9). In the energy range 20--200\,keV, we found that the pulsed fraction depends linearly on $\log E$ with slope $0.055\pm0.001$ (see Fig~\ref{pulsed}). The pulsed fraction increases significantly from 0.13 at 20 keV up to 0.18 at 150 keV, i.e., by about 30\%. 

To relate the measured pulsed fraction to the predicted position of the centroid, we simulated $10^6$ photons emanating from a point source at the position of the pulsar and an extended source with the extension of the Crab nebula, with a centroid shifted by 17.3" as found from \cra\ data. We convolved the resulting photon distribution with the large beam of ISGRI and fitted the distribution with a Gaussian to determine the position of the centroid. Unsurprisingly, our simulations indicate that the expected centroid depends linearly on the pulsed fraction. Therefore, using Fig. \ref{pulsed} the pulsed fraction can be directly compared to the observed shift. 

\subsection{Relative positions}
\label{relapos}

\begin{figure}
\resizebox{\hsize}{!}{\includegraphics{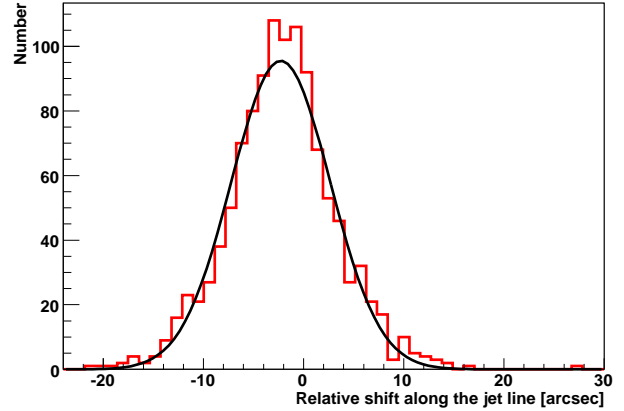}}
\caption{Distribution of the difference between the on- and off-pulse position measurements in each individual pointing, measured along the jet line. The distribution is significantly shifted by $2.28\pm0.15$ arcsec.}
\label{differences}
\end{figure}

In addition to the measurement of the centroid independently in the different bands and pulse phases, it is also possible to measure the difference between several parameters in each pointing individually. In this case, the systematic error of 0.5 arcsec, which probably originates in uncertainties in the spacecraft attitude reconstruction, cancels out, so it is more accurate to calculate the difference between 2 quantities in each pointing and calculate the mean rather than compute the mean of each 2 quantities and then their difference. This allows us to measure the difference between the positions during the on- and off-pulse phases with very good accuracy. Moreover, we have seen in Sect. \ref{interp} that all the positions lie along the jet line, so it is convenient to compute the difference between 2 quantities along this line. Figure \ref{differences} shows the distribution of the difference between the on- and off-pulse emission in the 20-40 keV band for each ScW, measured along the jet line. Fitting the distribution with a Gaussian, we find a very significant shift ($15\sigma$) of $2.28\pm0.15$ arcsec between the on- and off-pulse phases. The standard deviation of the distribution is found to be smaller than that of the 2 distributions individually, which proves that the systematic effects were at least partially cancelled. Perpendicular to the jet line, no shift is found ($\Delta=0.16\pm0.15$). This confirms that the centroid is moving along the jet line. A similar analysis can also be performed for the positions measured in different energy bands. All the measured relative shifts are presented in Table \ref{tabrel}.

From Table \ref{tabrel}, we can see that the difference between the on- and off-pulse phases seems to increase with energy. Since the pulsed fraction increases by $\sim30\%$ (see Fig. \ref{pulsed}), this effect can be easily explained by the increasing pulsed emission. We can also see from Table \ref{tabrel} that both the on- and off-pulse emission shift towards the pulsar with energy. While this is perfectly expected for the on-pulse emission as the signal from the pulsar becomes stronger, this argument cannot be valid for the off-pulse emission, where by construction the pulsed emission is suppressed. Therefore, the relatively significant shift ($4.3\sigma$) observed between the off-pulse emission in the 20-40 and 40-100 keV bands must be caused by a change in the morphology of the nebula.

\begin{table}
\begin{tabular}{lcccc}
\hline
Quantities measured & Relative angular distance ["]\\
\hline
\hline
On-pulse/Off-pulse, 20-40 keV & $2.28\pm0.15$\\
On-pulse/Off-pulse, 40-100 keV & $2.58\pm0.19$\\
On-pulse/Off-pulse, 100-200 keV & $3.65\pm0.98$\\
Total, 40-100/20-40 keV & $1.50\pm0.15$\\
Total, 100-200/20-40 keV & $2.14\pm0.49$\\
On-pulse, 40-100/20-40 keV & $1.25\pm0.19$\\
On-pulse, 100-200/20-40 keV & $2.84\pm0.63$\\
Off-pulse, 40-100/20-40 keV & $1.05\pm0.24$\\
Off-pulse, 100-200/20-40 keV & $1.51\pm0.84$\\
\hline
\end{tabular}
\caption{Mean of the relative distance between two quantities, measured along the jet line (see text).}
\label{tabrel}
\end{table}

\section{Discussion}
\label{disc}

As it has been shown above, IBIS/ISGRI is capable of measuring the position of the X-ray centroid of the Crab pulsar/nebula with unprecedented accuracy ($<1"$) in the 20-200 keV band, for the total emission as well as for the phase-resolved cases. The best-fit positions all lie along a line joining the X-ray centroid of the nebula to the Crab pulsar. 

To obtain physical insight using these results, we used \cra\ images in the soft (0.5-2 keV) and hard (2-10 keV) bands to create a hardness ratio map (2-10/0.5-2). We extracted \cra/ACIS RMFs and ARFs for this observation and used these spectral responses to convert between the 2-10/0.5-2 hardness ratio and the corresponding photon index. We fixed the galactic absorption in the direction of the source to the value measured by \xmm\ ($n_H=3.45\times10^{21}\,\mbox{cm}^{-2}$, \citet{willingale}, in agreement with the 21cm value) and simulated \cra\ spectra for a wide range of photon indices (1.4-3.0) to compute the corresponding hardness ratio. We then fitted the simulated data points using a $3^{rd}$ order polynomial and used the best-fit function to convert the hardness ratio map into a photon-index map. Our resulting photon-index map agrees with that of \citet{mori}. Finally, assuming pure power-law spectra we used our photon index map together with the hard band image to extrapolate to higher energies. As a result, we obtained extrapolated high-resolution maps in the 20-40, 40-100, and 100-200 keV bands, for which we measured the centroid of the nebula. The centroid of the extrapolated maps can be compared directly to the off-pulse emission measured by IBIS/ISGRI and presented in Tab. \ref{tabpos}. To estimate the uncertainty in the extrapolated centroids, we simulated 1,000 extrapolated maps with soft X-ray flux and photon index normally distributed in each pixel, and we computed the distribution of the simulated centroids. The $1\sigma$ error is then given by the standard deviation of the distribution. The resulting $1\sigma$ error is very small, from 0.06" in the 20-40 keV band up to 0.15" in the 100-200 keV band.

Comparing the extrapolated centroids with the ISGRI off-pulse measurements, we find that the emission is significantly offset. In the 20-40 keV band, the ISGRI off-pulse emission is measured to be 4.5" away from the extrapolated position ($6.4\sigma$), in the direction of the pulsar. A similar difference (4.8", $6\sigma$) is found in the 40-100 keV band. Consistently, we have seen that the centroid of the off-pulse emission shifts towards the pulsar with increasing energy (see Sect. \ref{relapos}). This indicates that ISGRI probes different physical regimes than \cra. Only 2 different explanations of this effect can be invoked: either the centroid of the nebula is moving towards the pulsar, or the steady flux from the pulsar increases dramatically.

According to \citet{tennant}, the minimum soft X-ray flux of the pulsar is $<1\%$ of the peak flux. Moreover, the authors conclude that the origin of the emission is probably non-thermal, and is due to the same emission mechanism as the pulse, so the steady flux should vary in a similar way as the peak flux. To produce a shift of 4.5" in the 20-40 keV band compared to the predicted position, the steady flux from the pulsar should represent $\sim35\%$ of the total flux, i.e. around 3 orders of magnitude greater than the value measured by \cra\ in soft X-rays. Therefore, we can reasonably exclude the possibility that the detected shift would be caused by an increase in the persistent flux from the pulsar. As a result, we conclude that the shift between the predicted and the measured position of the Crab nebula is produced by a change in morphology of the nebula. 

In the Crab nebula, the most energetic electrons are expected to be produced at the wind termination shock, in the inner ring \citep{weisskopf}. With increasing distance from the termination shock, the cut-off energy in the electron spectrum decreases because of synchrotron cooling \citep[see Fig. 3 of][]{atoyan}. In X-rays, the nebula is $\sim2$ times smaller than in the optical, and therefore the synchrotron emission in the outer regions probably cuts off in the optical/UV. Following this picture, it is natural to expect that a similar effect is happening when we compare the hard X-ray/soft $\gamma$-ray band with the soft X-rays. In the 20-40 keV band, we see in Fig. \ref{chandra} that the centroid of the emission is found to be very close to the inner ring, which corresponds to the region where the most energetic electrons are produced. This indicates that in the outer regions of the X-ray image the spectrum cuts off in the X-ray/hard X-ray range. This interpretation is supported by the presence of the well-known spectral break at $\sim100$ keV in the total spectrum of the source \citep{jung}, which corresponds to an energy of $\sim10^{14}$ eV for the electrons. In this framework, the measurements of the centroid of the emission around the break energy with arcsecond accuracy imply that the spectral break is due to a cut-off in the electron distribution away from the termination shock.

\begin{figure}
\resizebox{\hsize}{!}{\includegraphics{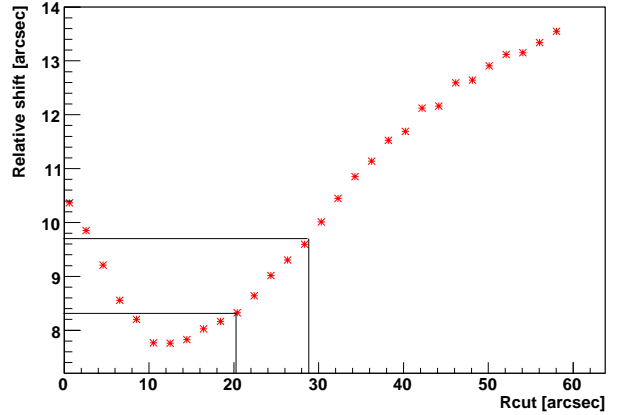}}
\caption{Relative shift of the centroid of the nebula with respect to the pulsar position in the 20-40 keV band as a function of the cut-off radius $R_{cut}$ (see text). The black lines show the acceptable range ($1\sigma$) for our measurement of the off-pulse centroid in the 20-40 keV band.}
\label{rcut}
\end{figure}

To validate this hypothesis, we extrapolated the \cra\ image to higher energies using this model rather than a pure power law. For simplicity, we assumed that the source is spherically symmetric around the pulsar and defined a radius $R_{cut}$ from the pulsar such that for $R<R_{cut}$ the emission is described by a pure power law, while for $R>R_{cut}$ an exponential cut-off is introduced. For the cut-off energy, we assumed a linearly-decreasing profile from $E_{cut}=40$ keV at $R=R_{cut}$ (the maximum of the corresponding ISGRI band) down to $E_{cut}=8$
keV in the outskirts (the upper limit of the \cra\ response). In physical terms, the characteristic synchrotron frequency of electrons with a Lorentz factor $\gamma$ in a magnetic field $B\sim0.3$ mG is given by (e.g. \citet{blumenthal})

\begin{equation}\nu_{syn}\sim7\times10^2\,\gamma^2\left(\frac{B}{0.3\mbox{ mG}}\right)\,\mbox{ Hz},\end{equation}

\noindent so we study electrons with a maximal energy $E\sim10^{13}-10^{14}$ eV. We then used this model to extrapolate the 2-10 keV \cra\ image to the 20-40 keV band and computed the centroid of the extrapolated image for a range of values of $R_{cut}$. Figure \ref{rcut} shows the predicted relative position of the centroid of the nebula with respect to the pulsar position, as a function of the cut-off radius $R_{cut}$. Surprisingly, for very small values of $R_{cut}$ the shift decreases with radius. This can be explained by the fact that for very small $R_{cut}$ a cut-off is present throughout the entire nebula, and therefore the morphology is not changing much. The function reaches a minimum at $R_{cut}\sim12$ arcsec, which corresponds to the radius at which the contrast between the central region (power law) and the outer parts (cut-off power law) is the greatest. For larger radius, the shift again increases as expected. 

In the 20-40 keV band, the off-pulse emission is shifted by $\sim$9 arcsec compared to the pulsar (see Tab. \ref{tabpos}). From Fig. \ref{rcut}, we see that 2 different values of $R_{cut}$ match with value, i.e., $R_{cut}\sim5$ arcsec and $R_{cut}\sim25$ arcsec. However, the smaller value corresponds to the very inner region of the nebula, and therefore a cut-off at this radius would have strong implications on the spectral shape, which is not seen, so we can safely reject the first solution. As a result, we find that the cut-off radius that corresponds to the observed centroid is $R_{cut}\sim25$ arcsec. At the distance of the Crab nebula ($d\sim2$ kpc, \citet{trimble}), this corresponds to a radius of $\sim0.35$ pc, i.e., to the outer regions of the torus. This proves that a cut-off effect in the outer regions of the nebula can reproduce the observed shift. Our modelling implies that at a radius $R\sim R_{cut}$, the synchrotron cut-off energy is $E\sim40$ keV, which corresponds to a maximal energy for the electrons of $\sim 6\times10^{13}$ eV.

Of course, these results are strongly model-dependent and our model is overly simplistic, in particular because of the assumption of spherical symmetry. Moreover, the output of our model depends on several input parameters, in particular on the choice of the maximum and minimum cut-off energy, so the derived values should be considered with care. However, the resulting numbers point towards a reasonable scenario for the electron distribution in the nebula. Indeed, our results are in excellent agreement with detailed theoretical modelling of the source using MHD simulations \citep{volpi,delzanna}, which predict that the jet and the external regions of the torus should disappear in hard X-rays/soft $\gamma$-rays \citep[see Fig. 4 of][]{volpi}. Therefore, our measurements qualitatively confirm the predictions of MHD simulations.

\section{Conclusion}

We have used the IBIS/ISGRI instrument on-board \intgr\ to measure the centroid of the Crab pulsar/nebula complex in hard X-rays/soft $\gamma$-rays with unprecedented accuracy, with the aim of investigating the behaviour of the relativistic electron population around the break energy ($E_{\mbox{\tiny{break}}}\sim100$ keV). Based on our understanding of the dependence of the astrometry of IBIS on the temperature of the mask, we have shown in Sect. \ref{syst} that despite its moderate angular resolution (12 arcmin FWHM), in the case of sources detected with high $S/N$ IBIS/ISGRI can measure the position of astrophysical sources with an accuracy $\sim0.5$ arcsec. Applying this method to the Crab, we found that the emission is offset significantly with respect to the position of the pulsar, by $\sim8"$ in the 20-40 keV band. We were also able to measure the position of the on- and off-pulse phases independently. All the measured positions are found along the line connecting the pulsar to the centroid of the X-ray emission in the torus, which corresponds roughly to the jet line (see Sect. \ref{interp}).

Performing phase-resolved imaging (see Sect. \ref{relapos}), we found that the on-pulse emission (where the Crab pulsar accounts for an important fraction of the flux) depends on energy: the on-pulse shift with respect to the pulsar position decreases by $\sim30\%$ between 20 and 200 keV. This effect can be easily explained by an increase in the pulsed fraction (see Sect. \ref{pulsfrac}).

More interestingly, we find that during the off-pulse phase (where the emission from the pulsar is negligible) the centroid of the source is significantly offset from the position predicted from \cra\ data (see Sect. \ref{disc}). This indicates that the morphology of the source changes around the break energy. Consistently, we show that a cut-off in the electron spectrum in the outer regions of the X-ray nebula, caused by synchrotron cooling and in agreement with the predictions of \citet{atoyan}, can reproduce the observed shift.

The centroid of the nebular emission around the break energy coincides with the inner ring, which is interpreted as the location to be the wind termination shock \citep{weisskopf}. Therefore, our data imply that above the break energy only the strong shock regions are responsible for the emission. Comparing our results with the predictions of a simple model for the spectral evolution of the nebula, we find that the electrons in the outer regions of the torus ($d\gtrsim0.35$ pc away from the pulsar) probably reach a maximal energy close to $10^{14}$ eV. This result agrees with theoretical studies carried out using MHD simulations \citep{volpi}. In the near future, because of a higher angular resolution by a factor $\sim20$ \emph{NuSTAR} will be able to resolve the Crab pulsar/nebula complex up to $\sim80$ keV, which will allow us to probe the electron spectrum in the different regions of the nebula close to the break energy.

\begin{acknowledgements}
We thank Bruce O'Neel for the integration of our thermal misalignment correction into the OSA software. DE thanks Sonja Hadj-Salem for her help. CF is supported by the German grant DLR~50~OG~0601. This work is based on observations with INTEGRAL, an ESA project with instruments and science data center funded by ESA member states (especially the PI countries: Denmark, France, Germany, Italy, Switzerland, Spain), Czech Republic and Poland, and with the participation of Russia and the USA.
\end{acknowledgements}

\bibliographystyle{aa}
\bibliography{crab}

\end{document}